\documentclass[times]{nmeauth}

\usepackage{moreverb}

\usepackage[dvips,colorlinks,bookmarksopen,bookmarksnumbered,citecolor=red,urlcolor=red]{hyperref}

\newcommand\BibTeX{{\rmfamily B\kern-.05em \textsc{i\kern-.025em b}\kern-.08em
T\kern-.1667em\lower.7ex\hbox{E}\kern-.125emX}}

\def\volumeyear{2022}

\usepackage{kurz}
\usepackage{amssymb}
\usepackage{mathrsfs}
\usepackage{multirow}    
\usepackage{amsmath}
\usepackage{caption}  
\usepackage{stmaryrd}
\usepackage{rotating}
\usepackage{setspace}
\usepackage[numbers,sort&compress]{natbib}

\usepackage{alphalph}

\usepackage{listings}
\usepackage{xcolor}
\definecolor{customgreen}{rgb}{0,0.6,0}
\definecolor{customgray}{rgb}{0.5,0.5,0.5}
\definecolor{custommauve}{rgb}{0.6,0,0.8}

\makeatletter
\newalphalph{\fnsymbolwrap}[wrap]{\@fnsymbol}{}
\makeatother
 
\definecolor{codegreen}{rgb}{0,0.6,0}
\definecolor{codegray}{rgb}{0.5,0.5,0.5}
\definecolor{codepurple}{rgb}{0.58,0,0.82}
\definecolor{backcolour}{rgb}{0.95,0.95,0.92}

\lstdefinestyle{mystyle}{
backgroundcolor=\color{backcolour},   
commentstyle=\color{codegreen},
keywordstyle=\color{magenta},
numberstyle=\tiny\color{codegray},
stringstyle=\color{codepurple},
basicstyle=\ttfamily\footnotesize,
breakatwhitespace=false,         
breaklines=true,                 
captionpos=t,                    
keepspaces=true,                 
numbers=left,                    
numbersep=5pt,                  
showspaces=false,                
showstringspaces=false,
showtabs=false,                  
tabsize=2
}

\definecolor{lightgreen}{rgb}{0.56, 0.93, 0.56}
\definecolor{moonstoneblue}{rgb}{0.45, 0.66, 0.76}

\begin{document}



\title{A simple hybrid linear and non-linear interpolation finite element for adaptive cracking elements method}

\author{Xueya Wang\affil{c,b}, Yiming Zhang\affil{a,b}\corrauth, Minjie Wen\affil{a}, Herbert Mang\affil{d,e,f}\corrauth}


\address{ 	\affilnum{a}School of Civil Engineering and Architecture, Zhejiang Sci-Tech University,  No. 2 Street, 310018~Hangzhou,~Zhejiang Province,~China\break 
	\affilnum{b}Jinyun Institute, Zhejiang Sci-Tech University,  WenYuXi Road, 321400~LiShui,~Zhejiang Province,~China\break 
	\affilnum{c}Faculty of Mechanical Engineering \& Mechanics, Ningbo University, Fenghua Road 818, 315211~Ningbo,~Zhejiang Province,~China \break	
	\affilnum{d}Institute for Mechanics of Materials and Structures (IMWS), Vienna University of Technology,Karlsplatz 13/202, 1040~Vienna, Austria \break
		\affilnum{e}State Key Laboratory for Disaster Reduction in Civil Engineering, Tongji University, Siping Road 1239, 200092~Shanghai,~China \break
		\affilnum{f}Department of Geotechnical Engineering, Tongji University, Siping Road 1239, 200092~Shanghai,~China 
}

\corraddr{Herbert A. Mang and Yiming Zhang\\
	\mbox{~~~~~~~~~~~~~~~~~~~~~~~Email:} Herbert.Mang@tuwien.ac.at, Yiming.Zhang@hebut.edu.cn}

\begin{abstract}
\Large
Cracking Elements Method (CEM) is a numerical tool to simulate quasi-brittle fractures, which does not need remeshing, nodal enrichment, or complicated crack tracking strategy.  The cracking elements used in the CEM can be considered as a special type of finite element implemented in the standard finite element frameworks.  One disadvantage of CEM is that it uses nonlinear interpolation of the displacement field (Q8 or T6 elements), introducing more nodes and consequent computing efforts than the cases with elements using linear interpolation of the displacement field.  Aiming at solving this problem, we propose a simple hybrid linear and non-linear interpolation finite element for adaptive cracking elements method in this work.  A simple strategy is proposed for treating the elements with $p$ edge nodes $p\in\left[0,n\right]$ and $n$ being the edge number of the element.  Only a few codes are needed.  Then, by only adding edge and center nodes on the elements experiencing cracking and keeping linear interpolation of the displacement field for the elements outside the cracking domain, the number of total nodes was reduced almost to half of the case using the conventional cracking elements.  Numerical investigations prove that the new approach inherits all the advantages of CEM with greatly improved computing efficiency.  

\end{abstract}


\maketitle

\vspace{-6pt}

\Large

\section{Introduction}
\label{sec:it}
In last decades, many novel numerical methods were invented to capture the strong discontinuity processes of quasi brittle materials.  Some models were built in discrete frameworks such as particle/mesh-free methods \cite{Rabczuk2010,Rabczuk2007473,Rabczuk20102437,Zhuang2011}.  Some other models were built in equivalent frameworks which introduce springs, lattices, and bonds to model the continuum media experiencing cracking \cite{Ren2017,Zhaogaofeng2017,Yuhaitao2020,Yiming:22}.  Despite of the advantages of these model simulating the discrete processes of continuum media such as cracking, they are more or less suffered from boundary effects, parameter calibration, or low computing efficiency induced by nonlocal effects.  On the other hand, because finite element method (FEM) is the most widely used numerical framework for continuum mechanics, the models built in FEM framework which are capable of capturing discontinuities are very preferable for scientists and engineers.  

Two basic routines to model discontinuities in the FEM framework are i) moving boundaries and ii) smeared damage \cite{DeBorst:02,Cervera:04,Cervenka:01}.  Moving boundary refers to the methods using remeshing and explicitly describing cracks \cite{Areias2009,Areias2013} and smeared damage refers to the methods applying continuous damage mechanics \cite{Cervera:10}.  The former necessitate complex topology algorithm and equivalent projections of the physical field, while the latter commonly suffer from mesh dependency.  In the recent two decades, enriched models were proposed which use original discretization and show weak mesh dependency, including nodal enrichment and elemental enrichment.  Nodal enrichment models introduce extra freedom degrees at nodes for describing the topology information of cracks such as eXtended Finite Element methods (XFEM) \cite{WuJianying:01,Belytschko:06}, phantom node methods \cite{HansboHansbo,SongandBelytschko} and Numerical Manifold Methods (NMM) \cite{Wuzhijun2012,ZhengHong:04,ZhengHong:05,WU201894}.  Many nodal enrichment methods can accurately capture the information around crack tips which meanwhile need complicate nodal enrichment functions and consequently extra freedom degrees.  Elemental enrichment models\footnote{In the following parts of this work, we use SDA (Strong Discontinuity embedded Appraoch) as the abbreviation of the elemental enrichment model though some researchers suggest to use E-FEM} focus on the localization/condensation\cite{Cervera2020,Mosler:01} procedures which attempt to digest the crack openings in local elements \cite{Oliver:11,Saloustros:01,Dias-da-Costa:02,NIKOLIC2018480}.  Different from nodal enrichment methods, most SDAs abandon accurate description of the crack tips but obtain higher efficiency and great flexibility \cite{Marasca2018,Borja:01}, indicating that SDAs are easier than nodal enrichment models to implement into conventional FEM framework.

After eliminating the stress locking problem of the SDAs with the standard statically optimal
symmetric formulation (SOS) \cite{Yiming:15}, we improved the original model and proposed a method named Cracking Element (CE) \cite{Yiming:15,Yiming:16,Yiming:19}, which uses piecewise cracking segments for representing continuous crack paths and does not necessitate cracking tracking strategies like most SDAs \cite{Saloustros2018,Cervera:03,Dumstorff:01}.  Furthermore, we presented a novel framework named Global Cracking Element (GCE) in \cite{Yiming:20}, where we proved that a special type of element can be built based on standard Galerkin finite elements.  By introducing a centre node for every cracked element the freedom degree of which is taken to represent crack openings.  GCE has been successfully adopted in different types of element \cite{Yiming:21} with different iteration procedures \cite{Yiming:23}, showing robustness and reliability.  However, GCE has a great shortcoming.  As proven in \cite{Yiming:15}, nonlinear interpolation of the displacement field is necessary for the cracking element to avoid stress locking.  Since it is unknown which elements will experience cracking in the beginning, higher order nonlinear elements shall be used to discretize the whole domain, bringing great computing efforts.

Aiming at solving this problem, we proposed a simple hybrid linear and non-linear interpolation finite element originally designed for adaptive cracking elements.  The main features of the new model are as follows:
\begin{itemize}
	\item
	A simple computing routine is proposed.  By only modifying the $\mathbf{N}$ vector and $\mathbf{B}$ matrix of the original nonlinear elements based on the position of edge nodes, the elemental stiffness matrix and residual vector of elements with $p$ edge nodes $p\in\left[0,n\right]$ where $n$ is the edge number of the element can be easily obtained.  The computing routine is so simple that only a few codes is needed (also provided).
	\item
	The domain is descretized into linear elements in the beginning.  Extra nodes will be adaptively introduced on the edges and centres of the elements in the cracking region for making them as higher order nonlinear elements and cracking elements.  Since cracking regions are relatively small comparing to the whole domain, computing efforts are reduced into almost half of the original global cracking elements model.
	\item 
	All the advantages of the global cracking elements are inherited, including no remeshing, no crack tracking, weak mesh dependency, standard finite element framework.
	\item 
	This strategy is suitable not only for cracking elements framework but also for all numerical methods potentially benefiting from discretization by hybrid elements with linear and nonlinear interpolations of the displacement field.  For example: adding edge nodes on the curved boundaries to enhance the precision.
\end{itemize}

The remainder of this paper is organized as follows: In Section~\ref{sec:ace}, the cracking element will be briefly introduced including the constitutive law and its elemental formulation (elemental $\mathbf{B}$, $\mathbf{K}$ matrices and residual vector).  Then the self-propagating crack procedure will be discussed.  Last but not least, the simple strategy for modelling elements with $p$ edge nodes $p\in\left[0,n\right]$ will be presented in details.  In Section~\ref{sec:NEs}, numerical examples are given to demonstrate the robustness and high efficiency of the approach.  Finally, Section~\ref{sec:conc} contains concluding remarks.

\section{Adaptive cracking elements}
\label{sec:ace}
\subsection{Characteristic length of cracking elements}
For quasi-brittle materials, the damaging zone is of limit width almost to zero.  Displacement jumps or crack openings (distance between the two crack surfaces, with dimension ``length") are used to quantitatively evaluate the seriousness of cracks.  On the other hand, in the framework of continuum mechanics, strain (dimensionless) is used to evaluate the seriousness of deformations.  Characteristic length \cite{Oliver:02} is a parameter to bridge these two quantities with which it is possible to model cracks in continuum mechanics, as used in classical smeared crack model \cite{Cervera:10}.

\begin{figure}[htbp]
	\centering
	\includegraphics[width=0.9\textwidth]{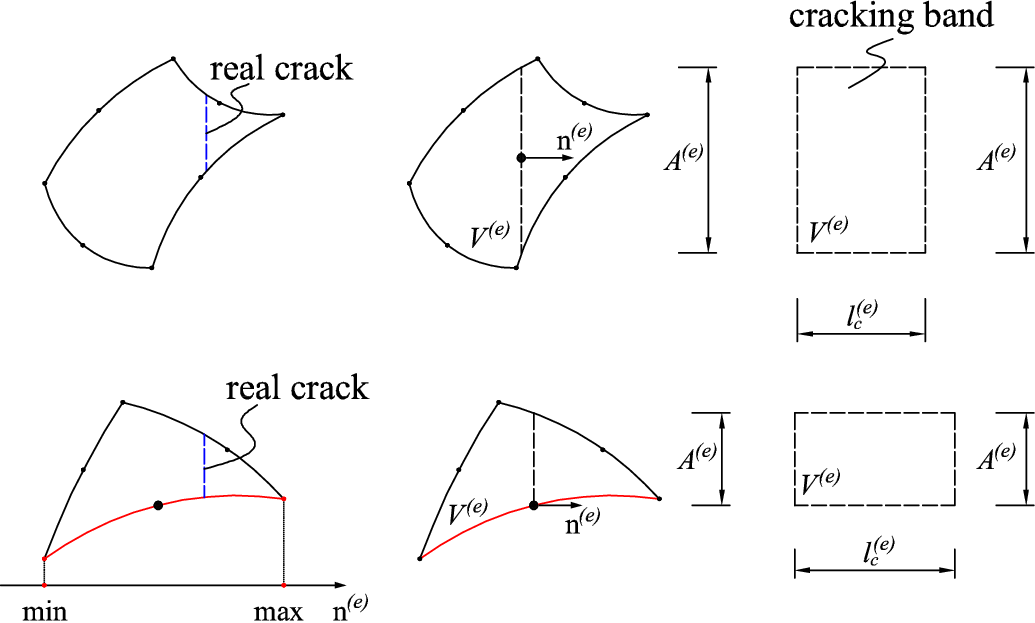}
	\caption{Relationships among $l_c$, $V^{(e)}$ and $A^{(e)}$ of element $e$ (based on a parallel crack passing through the center point of quadrilateral or triangular types of elements)}
	\label{fig:lc}
\end{figure}

In \cite{Yiming:11}, based on conservation of dissipated energy we proved that the characteristic length of cracking elements for quasi-brittle material depends only on the size and shape of elements but not materials.  For element ``$e$", its characteristic length $\l_c^{(e)}$ equals to $V^{(e)}\ / \ A^{(e)}$ where $V^{(e)}$ is the volume of ``$(e)$" and $A^{(e)}$ is the area of crack surface passing through the center point of quadrilateral type elements \cite{Yiming:14} or the mid point of one edge of triangular type of elements \cite{Yiming:21}.  The characteristic length helps to bridge the element experiencing cracking to a cracking band with a width equal to $l_c$, see Figure~\ref{fig:lc} for 2D condition, where $\mathbf{n}^{(e)}$ is the normal unit vector of crack.

\subsection{The constitutive relation}
\begin{figure}[htbp]
	\centering
	\includegraphics[width=0.55\textwidth]{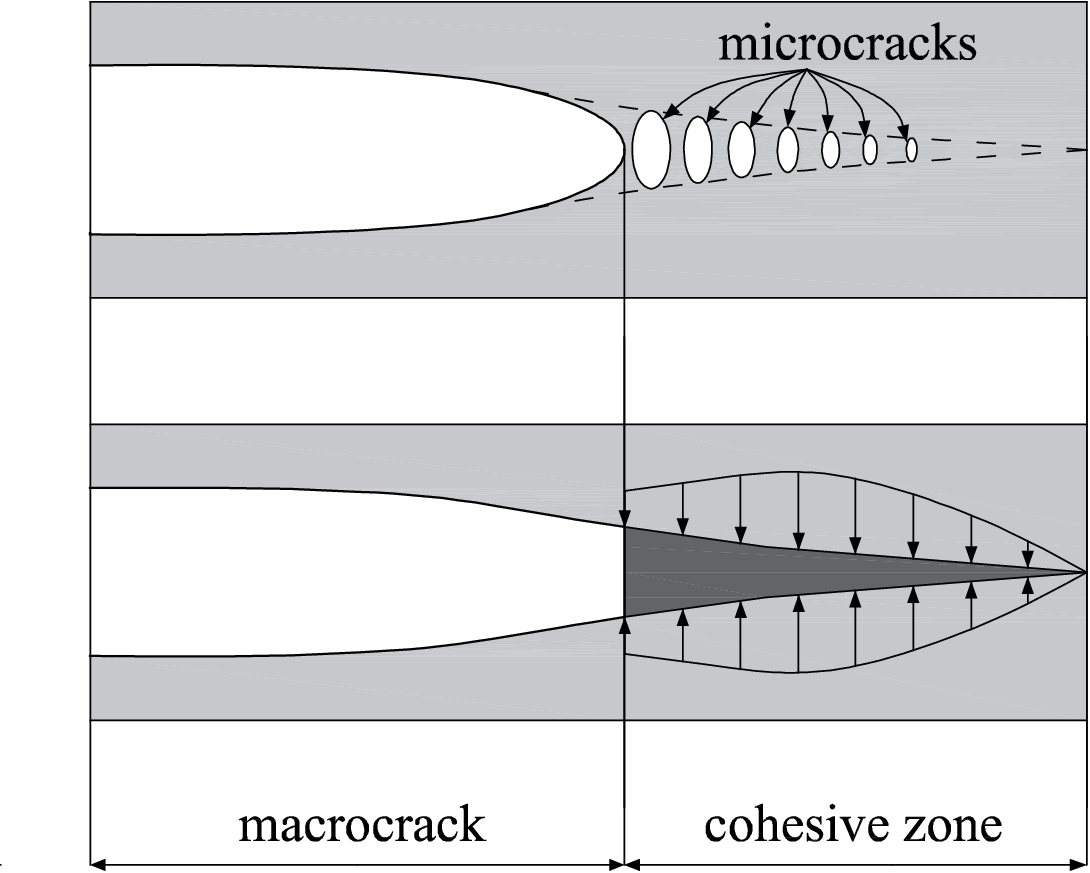}
	\caption{The cohesive zone model (redrawn based on the figure provided in \cite{Fernando2019})}
	\label{fig:cohesivezone}
\end{figure}

\begin{figure}[htbp]
	\centering
	\includegraphics[width=0.7\textwidth]{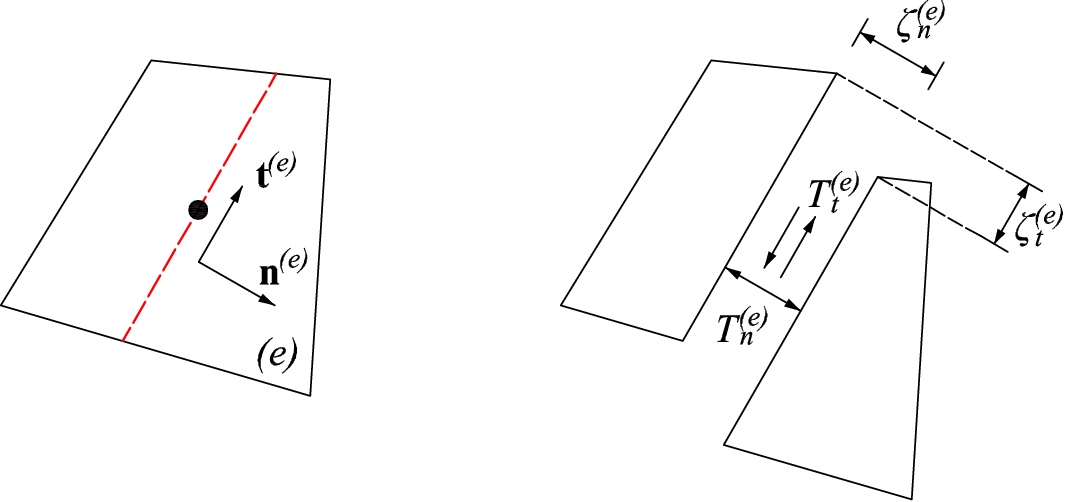}
	\caption{Unit vectors $\mathbf{n}^{(e)}$, $\mathbf{t}^{(e)}$, crack openings $\zeta_n^{(e)}$, $\zeta_t^{(e)}$, and traction forces $T_n^{(e)}$, $T_t^{(e)}$}
	\label{fig:CRACKele}
\end{figure}

We assume there is a crack in element ``$e$" whose normal and tangential unit vectors are known as $\mathbf{n}^{(e)}$ and $\mathbf{t}^{(e)}$ respectively.  In 2D conditions, $\mathbf{n}^{(e)}=\left[n^{(e)}_x,n^{(e)}_y\right]$ and $\mathbf{t}^{(e)}=\left[t^{(e)}_x,t^{(e)}_y\right]$.  Clearly, $\mathbf{n^{(e)}}\cdot\mathbf{t^{(e)}}=0$.  The crack opening along $\mathbf{n}^{(e)}$ and $\mathbf{t}^{(e)}$ are $\zeta_n^{(e)}$ and $\zeta_t^{(e)}$ respectively.  Considering the cohesive zone model (see Figure~\ref{fig:cohesivezone}), the traction forces between the crack opening are $T_n^{(e)}$ and $T_t^{(e)}$, see Figure~\ref{fig:CRACKele}.

\begin{figure}[htbp]
	\centering
	\includegraphics[width=0.7\textwidth]{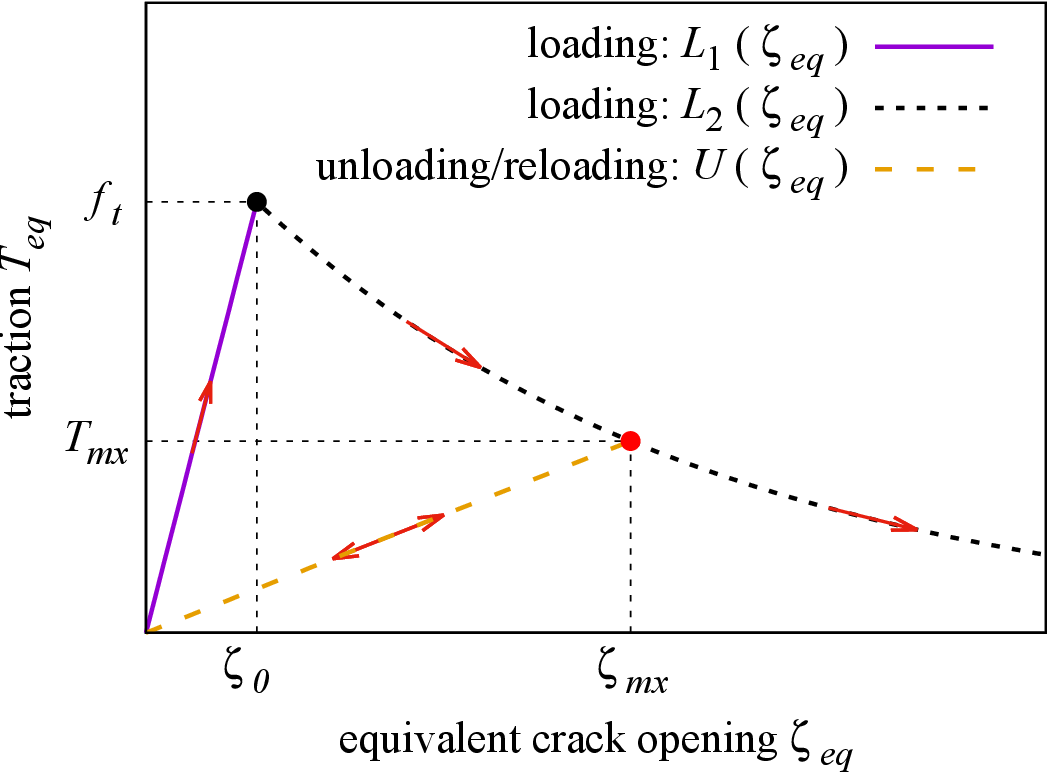}
	\caption{Traction-separation curve $T^{(e)}_{eq}\left(\zeta^{(e)}_{eq} \right)$}
	\label{fig:Traction}
\end{figure}

When using an exponential mixed-mode traction-separation law \cite{Meschke:01,Belytschko:02}, equivalent crack opening $\zeta^{(e)}_{eq}$ is introduced as 
\begin{equation}
\zeta^{(e)}_{eq}=\sqrt{\left(\zeta^{(e)}_n\right)^2+\left(\zeta^{(e)}_t\right)^2}.
\label{eq:zetaeq}
\end{equation}
Then we can obtain the traction forces as 
\begin{equation}
\begin{aligned}
&T^{(e)}_n=T^{(e)}_{eq}\frac{\zeta^{(e)}_n}{\zeta^{(e)}_{eq}},~ T^{(e)}_t=T^{(e)}_{eq}\frac{\zeta^{(e)}_t}{\zeta^{(e)}_{eq}}\\
&\mbox{where }\\
&T^{(e)}_{eq}\left(\zeta^{(e)}_{eq} \right)=\left\{\begin{array}{ll}
L_1\left(\zeta^{(e)}_{eq} \right)=\cfrac{f^{(e)}_t}{\zeta^{(e)}_{0}}\ \zeta^{(e)}_{eq}& \mbox{for loading } \zeta^{(e)}_{eq}\leq \zeta^{(e)}_0,\\
L_2\left(\zeta^{(e)}_{eq} \right)=f^{(e)}_t \ \mbox{exp}\left[-\cfrac{f^{(e)}_t\left(\zeta^{(e)}_{eq}-\zeta^{(e)}_0\right)}{G^{(e)}_f-G^{(e)}_{f,0}}\right]& \mbox{for loading } \zeta^{(e)}_{eq}>\zeta^{(e)}_0,\\
U\left(\zeta^{(e)}_{eq} \right)=\cfrac{T^{(e)}_{mx}}{\zeta^{(e)}_{mx}}\ \zeta^{(e)}_{eq}& \mbox{for unloading/reloading},
\end{array}\right.
\end{aligned}
\label{eq:Traction}
\end{equation}
in which $f^{(e)}_t$ is the uniaxial tensile strength and $G^{(e)}_f$ is the fracture energy of the material of ``$e$".  $G^{(e)}_{f,0}$ is the threshold value of $G^{(e)}_f$, with $G^{(e)}_{f,0}=0.01\ G^{(e)}_f$ assumed.  $\zeta^{(e)}_0$ is the corresponding threshold opening, with $\zeta^{(e)}_0=2\  G^{(e)}_{f,0}/f^{(e)}_t$.  $\zeta^{(e)}_{mx}$ denotes the maximum opening that a crack has ever experienced.  Its value is updated at the end of every loading step if $\zeta^{(e)}_{mx}>\zeta^{(e)}_0$.  $T^{(e)}_{mx}=L_2\left(\zeta^{(e)}_{mx}\right) $ is the corresponding traction, see Figure~\ref{fig:Traction}.  More details can be found in \cite{Yiming:20}.

From equations~\ref{eq:zetaeq} and~\ref{eq:Traction}, the following relations can be obtained
\begin{equation}
\begin{aligned}
&\mathbf{D}^{(e)}=\left[\begin{array}{cc}
{\partial T^{(e)}_n}/{\partial \zeta^{(e)}_n}&{\partial T^{(e)}_n}/{\partial \zeta^{(e)}_t}\\
{\partial T^{(e)}_t}/{\partial \zeta^{(e)}_n}&{\partial T^{(e)}_t}/{\partial \zeta^{(e)}_t}\\
\end{array}\right]=\\
&\left\{\begin{array}{ll}
\cfrac{f^{(e)}_t}{\zeta^{(e)}_0}\left[\begin{array}{cc}
1&0\\
0&1\\
\end{array}\right],& \mbox{loading}, \zeta^{(e)}_{eq}\leq \zeta^{(e)}_0\\
\\
-\cfrac{T^{(e)}_{eq}}{\left(\zeta^{(e)}_{eq}\right)^2}\left[\begin{array}{cc}
\cfrac{\left(\zeta^{(e)}_n\right)^2}{\zeta^{(e)}_{eq}}+\cfrac{f^{(e)}_t\ \left(\zeta^{(e)}_n\right)^2}{{G^{(e)}_f-G^{(e)}_{f,0}}}-\zeta^{(e)}_{eq}&
\cfrac{\zeta^{(e)}_n\ \zeta^{(e)}_t}{\zeta^{(e)}_{eq}}+\cfrac{f^{(e)}_t\ \zeta^{(e)}_n\ \zeta^{(e)}_t}{{G^{(e)}_f-G^{(e)}_{f,0}}}\\
\cfrac{\zeta^{(e)}_n\ \zeta^{(e)}_t}{\zeta^{(e)}_{eq}}+\cfrac{f^{(e)}_t\ \zeta^{(e)}_n\ \zeta^{(e)}_t}{{G^{(e)}_f-G^{(e)}_{f,0}}}&
\cfrac{\left(\zeta^{(e)}_t\right)^2}{\zeta^{(e)}_{eq}}+\cfrac{f^{(e)}_t\ \left(\zeta^{(e)}_t\right)^2}{{G^{(e)}_f-G^{(e)}_{f,0}}}-\zeta^{(e)}_{eq}\\
\end{array}\right],& \mbox{loading}, \zeta^{(e)}_{eq}>\zeta^{(e)}_0\\
\\
\cfrac{T^{(e)}_{mx}}{\zeta^{(e)}_{mx}}\left[\begin{array}{cc}
1&0\\
0&1\\
\end{array}\right],& \mbox{unloading/reloading},
\end{array}\right.
\label{eq:dTraction}
\end{aligned}
\end{equation}
which will be used for computing the stiffness matrix and the residual of the cracking element.
\subsection{The cracking elements}
\subsubsection{Enhanced strain and balance relation}
As presented in \cite{Yiming:20}, for a domain $\mathbf{x}$ experiencing cracking, its total strain ${\boldsymbol{\varepsilon}}$ consists of the elastic strain $\bar{\boldsymbol{\varepsilon}}$ and the enhanced strain $\tilde{\boldsymbol{\varepsilon}}$ as 
\begin{equation}
\begin{array}{ccccc}
\underbrace{\widehat{\boldsymbol{\varepsilon}}(\mathbf{x})=\nabla^S \bar{\mathbf{u}}(\mathbf{x})}&=  &\underbrace{\bar{\boldsymbol{\varepsilon}}(\mathbf{x})}  &+&\underbrace{\left[(\mathbf{n} \otimes \nabla \varphi)^S \zeta_n(\mathbf{x})+(\mathbf{t} \otimes \nabla \varphi)^S \zeta_t(\mathbf{x})\right]},\\
\mbox{total strain}&&\mbox{elastic strain}&&\mbox{enhanced strain } \widetilde{\boldsymbol{\varepsilon}}
\end{array}
\label{eq:EAS}
\end{equation}
where $\varphi(\mathbf{x})$ is a differentiable function with $\varphi(\mathbf{x})\in \left[0,1\right]$ and $\nabla \varphi$ is a vector with the dimension ``length$^{-1}$".  When considering the domain $\mathbf{x}$ is limited in element ``e", Eq.~\ref{eq:EAS} is obtained as
\begin{equation}
\begin{array}{cccc}
\bar{\boldsymbol{\varepsilon}}^{(e)}\approx&\underbrace{\sum^{n}_{i=1}\left(\nabla N^{(e)}_i \otimes \mathbf{u}_i\right)^S}&-&\underbrace{\frac{1}{\ l_c^{(e)} \ }\left[(\mathbf{n}^{(e)} \otimes  {\mathbf{n}}^{(e)}) \zeta_n^{(e)}+(\mathbf{n}^{(e)} \otimes  \mathbf{t}^{(e)})^S \zeta^{(e)}_t\right]},\\
&\widehat{\boldsymbol{\varepsilon}}^{(e)}&&\widetilde{\boldsymbol{\varepsilon}}^{(e)}
\end{array}
\label{eq:EAS_e}
\end{equation}
where $\left( \cdot \right)^S$ denotes the symmetric part of the tensor \cite{Mosler:01} and $\left( \cdot \right)^{(e)}$ is the value of the respective quantity of the element $e$.  E.g., $N_i^{(e)}$ is a shape function.  $n$ is the node number of element, in 2D condition, $n=8$ for quadratic element and $n=6$ for triangular element.

A local balance relation is built for solving extra freedom degree $\boldsymbol{\zeta}^{(e)}=\left[\zeta^{(e)}_n,\zeta^{(e)}_t\right]^T$ as
\begin{equation}
\begin{aligned}
&\boldsymbol{\sigma}^{(e)}=\mathbb{C}^{(e)}:\bar{\boldsymbol{\varepsilon}}^{(e)}\ \ \ \mbox{and}\\
&\left\{\begin{array}{l}
\left(\mathbf{n}^{(e)}\otimes\mathbf{n}^{(e)}\right):\boldsymbol{\sigma}^{(e)}-T_n^{(e)}=0\\
\left(\mathbf{n}^{(e)}\otimes\mathbf{t}^{(e)}\right)^S:\boldsymbol{\sigma}^{(e)}-T_t^{(e)}=0
\end{array}\right. ,
\label{eq:equcenter}
\end{aligned}
\end{equation}
where $\mathbb{C}^{(e)}$ is the elasticity tensor..Eq.~\ref{eq:equcenter} is obtained based on a simple rule: at center point of element, the traction forces of discrete form shall be equal to the cohesive forces of embedded form \cite{Yiming:11}.

For convenience, Voigt's notation is used for the remaining parts of this paper,  that $\boldsymbol{\varepsilon}^{(e)}$ and $\boldsymbol{\sigma}^{(e)}$ are written in matrix form as $\boldsymbol{\varepsilon}^{(e)}=\left[\varepsilon_x^{(e)},\varepsilon_y^{(e)},\gamma_{xy}^{(e)}\right]^T$ and $\boldsymbol{\sigma}^{(e)}=\left[\sigma_x^{(e)},\sigma_y^{(e)},\tau_{xy}^{(e)}\right]^T$.  

In \cite{Yiming:14}, a very important strategy named ``\emph{center representation}" is proposed that the strain/stress state at the center point of cracking element is used for representation of the stress/strain state of the whole element.  We assumed following matrices for further deduction.
\begin{equation}
\begin{aligned}
&\mathbf{B}^{(e),1}=
\left[ {\begin{array}{ccc}
	\mathbf{B}^{(e),1}_1&\cdots&\mathbf{B}^{(e),1}_n\\
	\end{array} } \right],\\
&\mbox{where}\\
&\mathbf{B}^{(e),1}_i=\left[\begin{array}{ccc}
\cfrac{\partial N_i^{(e),1}}{\partial x} &0\\
0&\cfrac{\partial N_i^{(e),1}}{\partial y} \\
\cfrac{\partial N_i^{(e),1}}{\partial y}&\cfrac{\partial N_i^{(e),1}}{\partial x}
\end{array}
\right],\ i=1 \cdots n,\\
\label{eq:B}
\end{aligned}
\end{equation} 
where $\left(\cdot\right)^{(e),1}$ denotes functions at the center point of $e$.  In other words, $N^{(e),1}$ is the shape function and $\mathbf{B}^{(e),1}$ is the $\mathbf{B}$ matrix at the origin of $\xi-\eta$ coordinate used by isoparametric element.  Then a special matrix $\mathbf{B}^{(e)}_\zeta$ is introduced as follows:
\begin{equation}
\mathbf{B}^{(e)}_\zeta=\frac{-1}{\ l_c^{(e)} \ }
\left[ {\begin{array}{c}
	\mathbf{n}^{(e)}\otimes\mathbf{n}^{(e)}\\
	\left(\mathbf{n}^{(e)}\otimes\mathbf{t}^{(e)}\right)^S
	\end{array} } \right]^T
=\frac{-1}{\ l_c^{(e)} \ }
\left[ {\begin{array}{cc}
	n^{(e)}_x\cdot n^{(e)}_x&n^{(e)}_x\cdot t^{(e)}_x\\
	n^{(e)}_y\cdot n^{(e)}_y&n^{(e)}_y\cdot t^{(e)}_y\\
	2\ n^{(e)}_x\cdot n^{(e)}_y&n_x\cdot t^{(e)}_y+n^{(e)}_y\cdot t^{(e)}_x
	\end{array} } \right].
\label{eq:Bz}
\end{equation} 

After assuming the displacement vector is given as $\mathbf{U}^{(e)}=\left[\mathbf{u}^{(e)}_1 \cdots \mathbf{u}^{(e)}_n\right]^T$.  Eq.~\ref{eq:EAS} becomes
\begin{equation}
\bar{\boldsymbol{\varepsilon}}^{(e)}=\bar{\boldsymbol{\varepsilon}}^{(e),1}=\left[ {\begin{array}{cc}
	\mathbf{B}^{(e),1}&	\mathbf{B}^{(e)}_\zeta
	\end{array} } \right] \left[ {\begin{array}{c}
	\mathbf{U}^{(e)}\\
	\boldsymbol{\zeta}^{(e)}
	\end{array} } \right].
\label{eq:EASQ8}
\end{equation}

Considering Eq.~\ref{eq:EASQ8}, Eq.~\ref{eq:equcenter} is rewritten as
\begin{equation}
\begin{aligned}
&\boldsymbol{\sigma}^{(e)}=\mathbf{C}^{(e)}\ \left[ {\begin{array}{cc}
	\mathbf{B}^{(e),1}&	\mathbf{B}^{(e)}_\zeta
	\end{array} } \right] \left[ {\begin{array}{c}
	\mathbf{U}^{(e)}\\
	\boldsymbol{\zeta}^{(e)}
	\end{array} } \right]\\
&\mbox{and}\\
&l_c^{(e)}\left(\mathbf{B}^{(e)}_\zeta\right)^T\mathbf{C}^{(e)}\ \left[ {\begin{array}{cc}
	\mathbf{B}^{(e),1}&	\mathbf{B}^{(e)}_\zeta
	\end{array} } \right]\left[ {\begin{array}{c}
	\mathbf{U}^{(e)}\\
	\boldsymbol{\zeta}^{(e)}
	\end{array} } \right]+\left[ {\begin{array}{c}
	T^{(e)}_n\\
	T^{(e)}_t
	\end{array} } \right]=\mathbf{0},
\label{eq:equcentermatrix}
\end{aligned}
\end{equation}
where $\mathbf{C}^{(e)}$ is the elastic matrix ($\mathbb{C}$ written in matrix form).  Eq.~\ref{eq:equcentermatrix} is a equation of $\boldsymbol{\zeta}^{(e)}$.  Meanwhile, consider the whole domain, momentum balance is 
\begin{equation}
\nabla \cdot \boldsymbol{\sigma} = \mathbf{F}
\label{eq:Mbalance}
\end{equation}
which is the same as its original form in continuum mechanics.

\subsubsection{Orientation of crack $\mathbf{n}^{(e)}$}
Cracking elements introduce an assumption for $\mathbf{n}^{(e)}$ that $\mathbf{n}^{(e)}$ is the first unit eigenvector of $\widehat{\boldsymbol{\varepsilon}}^{(e)}$ with $\widehat{\boldsymbol{\varepsilon}}^{(e)}=\mathbf{B}^{(e),1}\ \mathbf{U}^{(e)}$.

For 2D analysis, with
\begin{equation}
\widehat{\boldsymbol{\varepsilon}}^{(e)}\cdot\mathbf{n}^{(e)}-\widehat{{\varepsilon}}_1^{(e)}\cdot\mathbf{n}^{(e)}=\mathbf{0},
\label{eq:ep2D}
\end{equation}
$\widehat{{\varepsilon}}_1^{(e)}$ is obtained as 
\begin{equation}
\widehat{{\varepsilon}}_1^{(e)}=\frac{\widehat{{\varepsilon}}_x^{(e)}+\widehat{{\varepsilon}}_y^{(e)}+\sqrt{\left(\widehat{{\varepsilon}}_x^{(e)}-\widehat{{\varepsilon}}_y^{(e)} \right)^2+\left(\widehat{\gamma}_{xy}^{(e)}\right)^2}}{2}.
\label{eq:ep1}
\end{equation}
A great advantage of this assumption is that in a single iteration step $\mathbf{n}^{(e)}$ is a function of $\mathbf{U}^{(e)}$ but not $\boldsymbol{\zeta}^{(e)}$.  A disadvantage is $\mathbf{n}^{(e)}$ will slightly change in different steps.  The change of $\mathbf{n}^{(e)}$ reflects the change of orientation of the equivalent crack in local element caused by the redistribution of stresses.  With the propagation of cracks, $\mathbf{n}^{(e)}$ becomes steady.

Moreover, the uncracked elements of the domain are separated into two regions: the propagation region and the new root-search region.  This is done with a simple criterion: whether the element shares at least one boundary with a cracking element.  If the answer is ``yes", the element belongs to the propagation region.  Otherwise, it belongs to the new root-search region.  The following strategy is used to find the next cracking element:

\begin{equation}
\begin{aligned}
&\mbox{find }\ \mbox{max}\left\{\phi_{RK}^{(e)}\right\} \mbox{, with} \\
&\phi_{RK}^{(e)}=\left[ {\begin{array}{c}
	n^{(e)}_x\cdot n^{(e)}_x\\
	n^{(e)}_y\cdot n^{(e)}_y\\
	2\ n^{(e)}_x\cdot n^{(e)}_y
	\end{array} } \right]^T\ \mathbf{C}^{(e)}\ \widehat{\boldsymbol{\varepsilon}}^{(e)}-f_t^{(e)}\\
&\mbox{and }\phi_{RK}^{(e)}>0,
\label{eq:phiRK}
\end{aligned}
\end{equation}
where the propagation region is always at first searched.  If a new cracking element appears in the propagation region, this region will consequently expand, and the search will be continued.  If no new cracking element is found in the propagation region, the new root-search region will be checked.  If that a new cracking element appears in the new root-search region, the propagation region will further expand and the search in the propagation region will again be continued.  If still no new cracking element is found in the new root-search region, the next loading step will start, see \cite{Yiming:14,Yiming:20} for more details.

\subsubsection{The elemental stiffness matrix and residual vector}
Although cracking elements formulation is highly localized which is easy for coding in FEM framework and does not need crack tracking strategy.  Elements with nonlinear interpolation of displacement field are need for avoiding stress locking, see \cite{Yiming:11}.  In 2D analysis, 8 node quadrilateral element or 6 node triangular element are needed.  With global cracking elements \cite{Yiming:20,Yiming:21}, the numerical stability is greatly improved by introducing a center node in each cracking element, whose freedom degrees are crack openings $\zeta$.  Hence the elemental freedom degree of cracking elements are $\left[\mathbf{U}^{(e)},\boldsymbol{zeta}^{(e)}\right]^T$.  For convenience, we assume 
\begin{equation}
\boldsymbol{\mathsf{U}}=\bigwedge \left[ {\begin{array}{c}
	\mathbf{U}^{(e)}\\
	\boldsymbol{\zeta}^{(e)}
	\end{array} } \right],
\label{eq:assembleU}
\end{equation}
where $\bigwedge \left(\cdot\right)$ denotes the assemblage of the elemental matrix or vector in the global form.  Furthermore, in the remaining parts of the paper all subscripts and incremental symbol will act inside $\bigwedge$.  For example $\Delta \boldsymbol{\mathsf{U}}_i=\bigwedge \left[\Delta \mathbf{U}^{(e)}_i, \Delta \boldsymbol{\zeta}^{(e)}_i\right]^T$.

Considering standard Newton-Raphson (N-R) iteration procedure, for the iteration step $l$ within load step $i$ the following relation is obtained:
\begin{equation}
\begin{array}{cccc}
\boldsymbol{\mathsf{U}}_{i,l}=
&\underbrace{
	\boldsymbol{\mathsf{U}}_{i-1}+
	\Delta\boldsymbol{\mathsf{U}}_{l-1}}&+&\underbrace{
	\Delta\Delta\boldsymbol{\mathsf{U}}},\\
&\mbox{known}&&\mbox{unknown}\\
\end{array}
\label{eq:UdU2}
\end{equation}
where $\Delta\left(\cdot\right)$ denotes an increment with respect to the corresponding value at the preceding load step, $i-1$, and  $\Delta\Delta\left(\cdot\right)$ stands for an increment with respect to the value at the last N-R iteration step, $l-1$.  Then, the linearized balance equation is represented as
\begin{equation}
\boldsymbol{\mathsf{K}}_{l-1}\ \left( \Delta\Delta\boldsymbol{\mathsf{U}}\right)=\boldsymbol{\mathsf{R}}_{l-1},
\label{eq:linearizedBalance}
\end{equation}
where $\boldsymbol{\mathsf{K}}_{l-1}$ is the global stiffness matrix, with $\boldsymbol{\mathsf{K}}_{l-1}=\bigwedge \mathbf{K}^{(e)}_{l-1}$.  It can be found that $\left[\mathbf{B}^{(e)},\mathbf{B}_\zeta^{(e)}\right]$ and $\mathbf{K}_{l-1}^{(e)}$ are formally very similar to the $\mathbf{B}$ matrix and elemental stiffness $\mathbf{K}$ matrix of non-serendipity higher order elements i.e. Q9 and T7 elements in 2D condition.  This was firstly pointed out in \cite{Yiming:20}.

$\boldsymbol{\mathsf{R}}_{l-1}$ is the residual vector, with $\boldsymbol{\mathsf{R}}_{l-1}=\bigwedge \mathbf{R}^{(e)}_{l-1}$. $\boldsymbol{\mathsf{R}}_{l-1}$ is a function of $\left(\boldsymbol{\mathsf{U}}_{i-1}+\Delta\boldsymbol{\mathsf{U}}_{l-1}\right)$.

Similar to Eq.~\ref{eq:B}, we introduce
\begin{equation}
\begin{aligned}
&\mathbf{B}^{(e)}=
\left[ {\begin{array}{ccc}
	\mathbf{B}^{(e)}_1&\cdots&\mathbf{B}^{(e)}_n\\
	\end{array} } \right],\\
&\mbox{where}\\
&\mathbf{B}^{(e)}_i=\left[\begin{array}{ccc}
\cfrac{\partial N_i^{(e)}}{\partial x} &0\\
0&\cfrac{\partial N_i^{(e)}}{\partial y} \\
\cfrac{\partial N_i^{(e)}}{\partial y}&\cfrac{\partial N_i^{(e)}}{\partial x}
\end{array}
\right],\ i=1 \cdots n.\\
\label{eq:BGauss}
\end{aligned}
\end{equation} 
$\mathbf{B}^{(e)}$ has different values on different Gauss points.

Then, for the elemental stiffness matrix $\mathbf{K}^{(e)}$ and the residual vector $\mathbf{R}^{(e)}$ of the cracking element, we have
\begin{equation}
\mathbf{K}^{(e)}_{l-1}=\int  \left[ {\begin{array}{c}
	\mathbf{B}^{(e)}\\
	\mathbf{B}_\zeta^{(e)}
	\end{array} } \right]  \ \mathbf{C}^{(e)}\ \left[ {\begin{array}{cc}
	\mathbf{B}^{(e)}, \mathbf{B}_\zeta^{(e)}
	\end{array} } \right] \ d(e) +
\left[ {\begin{array}{cc}
	\mathbf{0}&\mathbf{0}\\
	\mathbf{0}&{A^{(e)}}\ \mathbf{D}^{(e)}\\
	\end{array} } \right],
\label{eq:crackedK}
\end{equation}
where $\int(\cdot)d(e)$ is the integral of function in element ``$e$".  $A^{(e)}$ is obtained based on Figure~\ref{fig:lc}.  $\mathbf{D}^{(e)}$ is obtained based on Eq.~\ref{eq:dTraction}.  Since $\mathbf{n}^{(e)},\boldsymbol{\zeta}^{(e)},l_c^{(e)}$ change at different iteration step $l$.  $\mathbf{K}^{(e)}_{l-1}$ shall be updated.
 $\mathbf{R}^{(e)}_{l-1}$, is obtained as
\begin{equation}
\mathbf{R}^{(e)}_{l-1}=\left[\begin{array}{c}
\mathbf{F}^{(e)}\\
-A^{(e)}\ \mathbf{T}^{(e)}
\end{array}\right]
-\mathbf{S}_{l-1}^{(e)}\
\left[ {\begin{array}{c}
	\mathbf{U}^{(e)}_{i-1}+\Delta \mathbf{U}^{(e)}_{l-1}\\
	\boldsymbol{\zeta}^{(e)}_{i-1}+\Delta \boldsymbol{\zeta}^{(e)}_{l-1}
	\end{array} } \right],
\label{eq:crackedR}
\end{equation}
where $\mathbf{T}^{(e)}=\left[T^{(e)}_n, T^{(e)}_t\right]^T$ (see Eq.~\ref{eq:Traction}) and $\mathbf{S}_{l-1}^{(e)}$ is a designed unsymmetrical matrix
\begin{equation}
\mathbf{S}_{l-1}^{(e)}=\left[ {\begin{array}{cc}
	\int \left(\mathbf{B}^{(e)}\right)^T \mathbf{C}^e \ \left(\mathbf{B}^{(e),1}\right) d(e) &\int \left(\mathbf{B}^{(e)}\right)^T \mathbf{C}^e \ \mathbf{B}_\zeta^{(e)} d(e) \\
	V^{(e)} \left(\mathbf{B}^{(e)}_\zeta\right)^T\mathbf{C}^e \ \left(\mathbf{B}^{(e),1}\right) &V^{(e)} \left(\mathbf{B}^{(e)}_\zeta\right)^T\mathbf{C}^e \ \mathbf{B}^{(e)}_\zeta\\
	\end{array} } \right].
\label{eq:crackedS}
\end{equation}
These equations are obtained from Eqs.~\ref{eq:equcentermatrix} and Eqs.~\ref{eq:Mbalance}.  The detailed deduction can be found in \cite{Yiming:20}.

\subsection{Uncracked elements with hanging nodes}
\begin{figure}[htbp]
	\centering
	\includegraphics[width=0.8\textwidth]{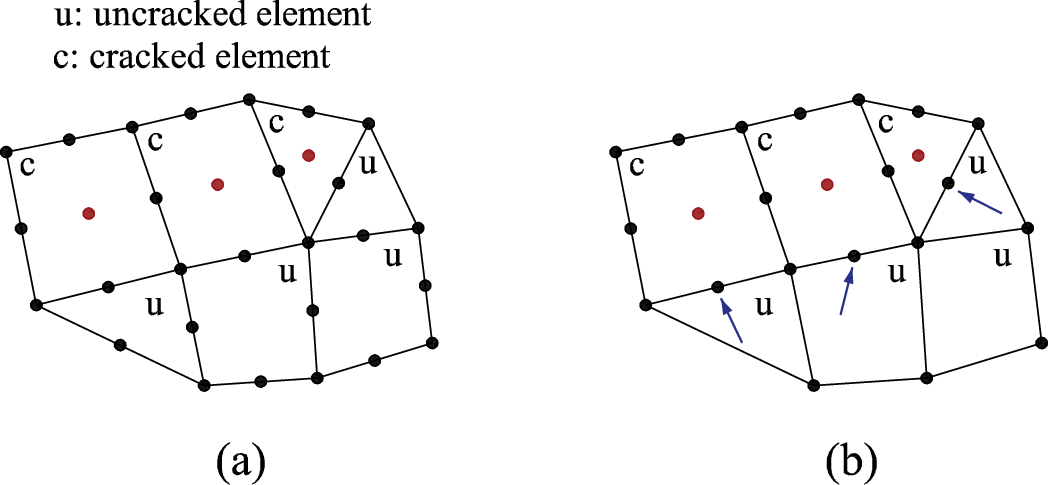}
	\caption{Classical and adaptive cracking elements: (a) classical cracking elements, (b) adaptive cracking elements and hanging nodes (see the blue arrows)}
	\label{fig:hanging}
\end{figure}

Taking 2D condition for example, in the classical version of cracking elements, the domain will be descretized by higher order elements i.e. Q8 and T6.  Then, once an element experiences cracking.  An extra node will be added at its center and its $\boldsymbol{\zeta}^{(e)}$ will become global unknown.  An obvious drawback of this strategy is that the cracking region mostly occupies only a few percentages of the whole domain whose position are commonly unknown in the beginning.  Great amount of higher order elements keep uncracked in the whole loading process and their higher order interpolations of displacement field bring limited benefits and affect efficiency.  Adaptive strategy will be: i) descretizing the domain with lower order elements in the beginning, ii) only introducing edge and center nodes for cracking elements in the calculation, see Figure~\ref{fig:hanging}.  The key point of this strategy is:``how to deal with elements with hanging nodes?" .

The history of hanging nodes is almost as long as that of FEM method.  Classical strategy is to introduce extra restraint on the hanging nodes\cite{BYFUT20172092}.  Sometimes advanced methods such as polygon finite elements\cite{WU2023935,NGUYENXUAN201745}, virtual elements\cite{Wriggers2020,Brezzi2021}, and scaled boundary elements\cite{YANG2007669,OOI2016154,CHEN201726} are implemented together.  On the other hand, this work only needs to deal with the quadrilateral elements with 4$\sim$8 nodes or triangluar elements with 3$\sim$6 nodes.  Comparing to the traditional methods focusing on the hanging nodes, this work focuses on the edge without hanging nodes, then presents a very simple formulation with very few coding efforts.

\begin{figure}[htbp]
	\centering
	\includegraphics[width=0.8\textwidth]{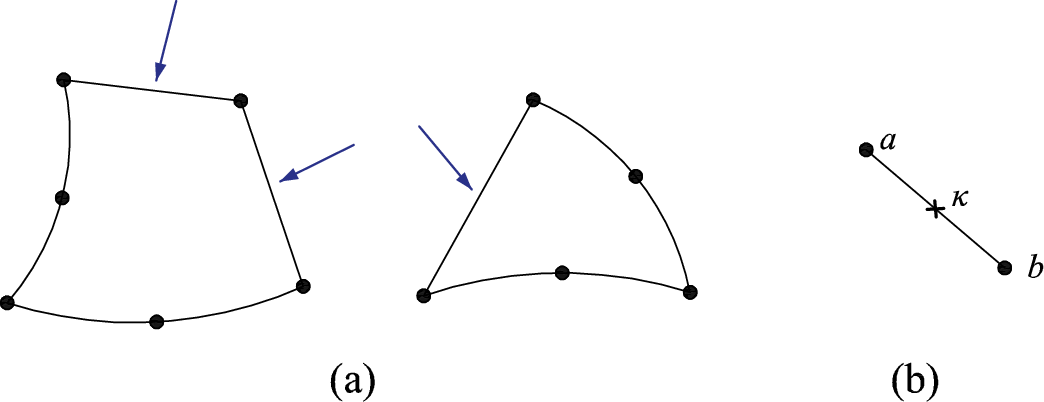}
	\caption{Elements with hanging nodes: (a) quadrilateral and triangluar elements with some hanging nodes, (b)an edge without hanging node}
	\label{fig:withouthanging}
\end{figure}

The elements with some hanging nodes are shown in Figure~\ref{fig:withouthanging}(a).  Considering an edge without hanging node connecting two nodes ``$a$" and ``$b$", we assume there is a virtual node ``$\kappa$" locating in the middle of $a$ and $b$, see Figure~\ref{fig:withouthanging}(b).  This process is conducted for all edges without hanging nodes.  Then we will obtain 8-node quadrilateral or 6-node triangluar elements with following shape function.
\begin{equation}
\begin{aligned}
&\mathbf{U}\left(\mathbf{x}\right)=\mathbf{N}\left(\mathbf{x}\right)\mathbf{U}^{(e)}\\
&\mbox{where}\\
&\mathbf{N}\left(\mathbf{x}\right)=\left[\cdots N_a\left(\mathbf{x}\right)\cdots  N_b\left(\mathbf{x}\right)\cdots N_\kappa\left(\mathbf{x}\right)\cdots \right]\\
&\mathbf{U}^{(e)}=\left[\cdots \mathbf{U}_a\cdots  \mathbf{U}_b\cdots \mathbf{U}_\kappa \cdots \right]^T.
\end{aligned}
\label{eq:shapehanging}
\end{equation}

\begin{figure}[htbp]
	\centering
	\includegraphics[width=0.8\textwidth]{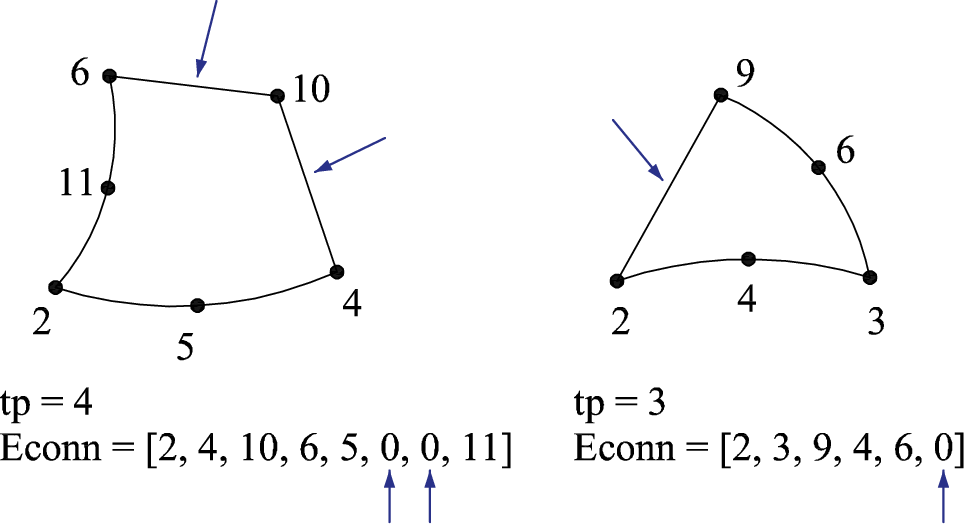}
	\caption{Elements with hanging nodes and their tp and Econn (the numbers indicate the id of nodes)}
	\label{fig:codeexplain}
\end{figure}

In the meanwhile, $\kappa$ locate in the middle of $a$ and $b$, indicating $\mathbf{U}_\kappa=0.5\left(\mathbf{U}_a+\mathbf{U}_b\right)$.  Hence, Eq.~\ref{eq:shapehanging} can be modified into
\begin{equation}
\begin{aligned}
&\mathbf{U}\left(\mathbf{x}\right)=\mathbf{N}\left(\mathbf{x}\right)\mathbf{U}^{(e)}\\
&\mbox{where}\\
&\mathbf{N}\left(\mathbf{x}\right)=\left[\cdots \left(N_a\left(\mathbf{x}\right)+\frac{N_\kappa\left(\mathbf{x}\right)}{2}\right)\cdots  \left(N_b\left(\mathbf{x}\right)+\frac{N_\kappa\left(\mathbf{x}\right)}{2}\right)\cdots 0\cdots \right]\\
&\mathbf{U}^{(e)}=\left[\cdots \mathbf{U}_a\cdots  \mathbf{U}_b\cdots \mathbf{0} \cdots \right]^T.
\end{aligned}
\label{eq:shapehanging2}
\end{equation}
Similarly, considering $\mathbf{B}^{(e)}$ matrix in Eq.~\ref{eq:BGauss}, we have following modified version
\begin{equation}
\mathbf{B}^{(e)}=\left[\cdots\left(\mathbf{B}^{(e)}_a+\frac{\mathbf{B}^{(e)}_\kappa}{2}\right)\cdots \left(\mathbf{B}^{(e)}_b+\frac{\mathbf{B}^{(e)}_\kappa}{2}\right)\cdots \mathbf{0}\cdots \right].
\label{eq:shapehanging2}
\end{equation}
\emph{When using Eq.~\ref{eq:shapehanging2} for building the elemental stiffness matrix and residual vector, the $\kappa$ row and $\kappa$ column will automatically become zero, which will not be assembled into the global stiffness matrix and residual vector (since $\kappa$ node does not exist).}  Hence, only $\mathbf{N}$ vector and $\mathbf{B}$ matrix needed to be modified in the code level.  A Fortran subroutine is provided as follows:
\begin{lstlisting}[language=fortran, caption=Fortran code: modifying $\mathbf{N}$ vector and $\mathbf{B}$ matrix, label=listing:process]
subroutine Process(tp,N,dNxy,Econn)
integer i,j,k,l,tp,Econn(*)
real*8 N(*),dNxy(2,*)
do i=tp+1,tp*2
   j=i-tp
   if(i.eq.tp*2) then
      k=1
   else
      k=i-tp+1
   endif
   if(Econn(i).eq.0) then
      N(j)=N(j)+0.5D0*N(i)
      N(k)=N(k)+0.5D0*N(i)
      N(i)=0.D0
      do l=1,2
         dNxy(l,j)=dNxy(l,j)+0.5D0*dNxy(l,i)
         dNxy(l,k)=dNxy(l,k)+0.5D0*dNxy(l,i)
         dNxy(l,i)=0.D0
      enddo            
   endif
enddo
end
\end{lstlisting}
where  ``N" is the $\mathbf{N}$ vector.  ``dNxy(1,*)" is $\partial \mathbf{N}\ / \ \partial x$ and ``dNxy(2,*)" is $\partial \mathbf{N}\ / \ \partial y$.  The original $\mathbf{N}$ and $\partial \mathbf{N}\ / \ \partial \mathbf{x}$ are obtained based on the 8-node quadrille and 6-node triangular elements, which will be modified by this subroutine.  ``tp" is a parameter controlling quadrille ($\mbox{tp}=4$) or triangular ($\mbox{tp}=3$) elements.  ``Econn" is an array recording the connection relation of nodes of the element.  Taking the elements shown in Figure~\ref{fig:withouthanging} for an example, their corresponding tp and Econn are provided in Figure~\ref{fig:codeexplain}.

In summary, for the uncracked elements, no matter how many hanging nodes they have, we use Eq.~\ref{eq:shapehanging2} and subroutine provided in Listing~\ref{listing:process} to obtain the modified $\mathbf{B}$ matrix then obtain following elemental stiffness matrix $\mathbf{K}^{(e)}_{l-1}$ and residual vector $\mathbf{R}^{(e)}_{l-1}$
\begin{equation}
\mathbf{K}^{(e)}_{l-1}=\int \left(\mathbf{B}^{(e)}\right)^T \ \mathbf{C}^{(e)}\ \mathbf{B}^{(e)} \ d(e)
\label{eq:uncrackedK},
\end{equation}
\begin{equation}
\mathbf{R}^{(e)}_{l-1}=\mathbf{F}^{(e)}-\mathbf{K}^{(e)}_{l-1}\left(\mathbf{U}^{(e)}_{i-1}+\Delta \mathbf{U}^{(e)}_{l-1}\right).
\label{eq:uncrackedR}
\end{equation}

Again, we would like to emphasize that $\mathbf{K}^{(e)}_{l-1}$ and $\mathbf{R}^{(e)}_{l-1}$ are formally the stiffness matrix and residual vector of 8-node quadrille (in 2D condition, $\mathbf{K}^{(e)}_{l-1}$ is with 16 rows and 16 columns and $\mathbf{R}^{(e)}_{l-1}$ is with 16 rows) and 6-node triangular elements (in 2D condition, $\mathbf{K}^{(e)}_{l-1}$ is with 12 rows and 12 columns and $\mathbf{R}^{(e)}_{l-1}$ is with 12 rows).  But the values at $\kappa$ rows and $\kappa$ columns (virtual nodes) are all zeros which will not be assembled into the global stiffness matrix and residual vector which have no position for these virtual nodes.

\section{Numerical examples}
\label{sec:NEs}
Plane-stress assumptions hold for all the numerical examples presented in this section.  
\subsection{L-shaped panel test}
\begin{figure}[htbp]
	\centering
	\includegraphics[width=0.99\textwidth]{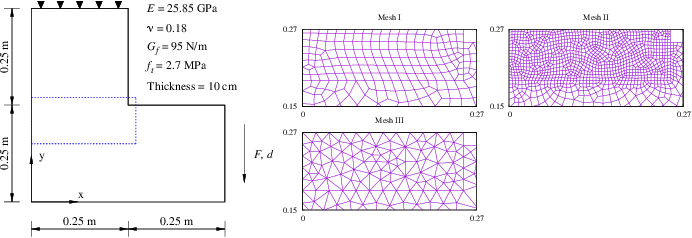}
	\caption{L-shaped panel test: the set-up, material parameters and meshes (unit: m)}
	\label{fig:LPModel}
\end{figure}

The L-shaped panel test, investigated in this paper, was studied before in \cite{Yiming:14,Yiming:20}.  The set-up, material parameters and meshes are shown in Figure~\ref{fig:LPModel}.  The displacement increment is chosen as $\Delta d=10\ \mu\mbox{m}$.

\begin{figure}[htbp]
	\centering
	\includegraphics[width=0.90\textwidth]{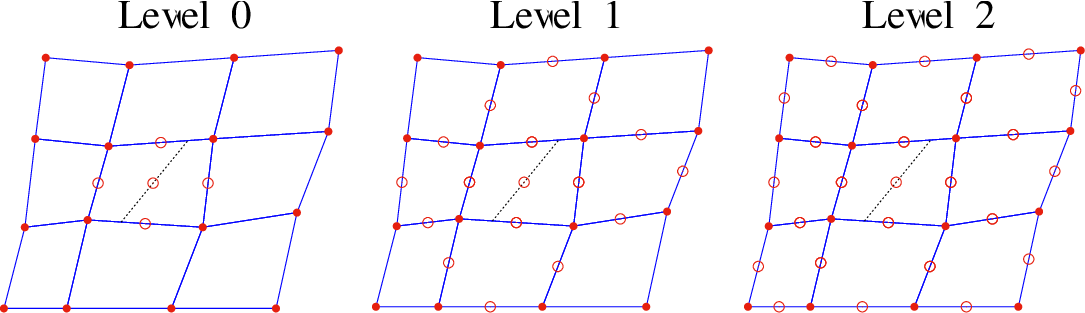}
	\caption{Three strategies for adding nodes}
	\label{fig:adaplevel}
\end{figure}

In this example, we compare three adaptive strategies in the simulation as
\begin{enumerate}
	\item 
	Level 0: extra nodes will be added only on the element experiencing cracking.
	\item
	Level 1: extra nodes will be added on the element experiencing cracking and surrounding elements sharing edges with the cracking element.
	\item 
	Level 2: extra nodes will be added on the element experiencing cracking and surrounding elements sharing at least one node with the cracking element,
\end{enumerate}
as shwon in Figure~\ref{fig:adaplevel}.

\begin{figure}[htbp]
	\centering
	\includegraphics[width=0.99\textwidth]{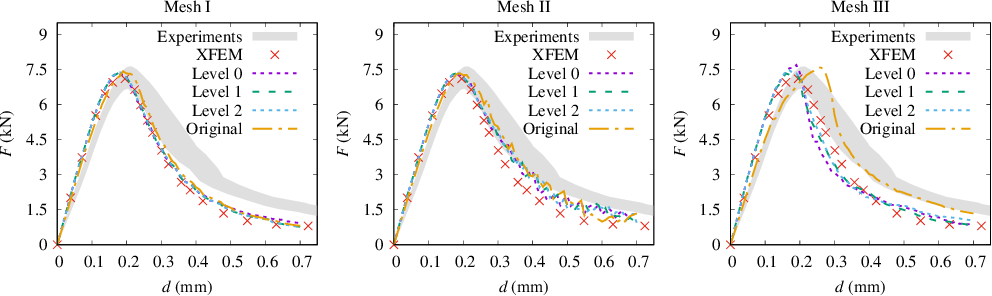}
	\caption{Force-displacement curves of the L-shaped panel for different meshes with different adaptive strategies, comparing the results for the GCEM to the experimental results \cite{Winkler:02} and the results obtained by the XFEM \cite{Meschke:01}}
	\label{fig:LPanelForce}
\end{figure}

\begin{figure}[htbp]
	\centering
	\includegraphics[width=0.99\textwidth]{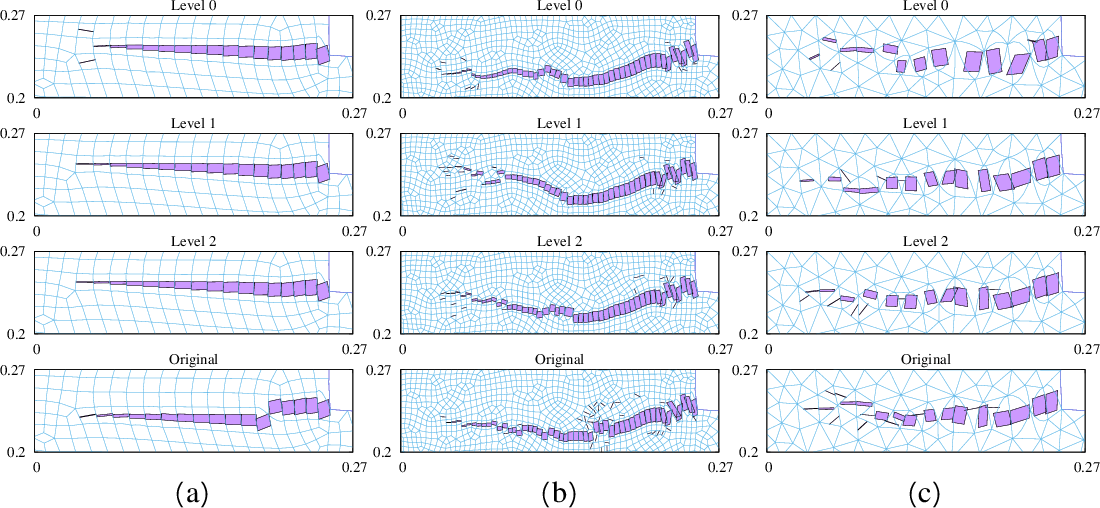}
	\caption{The crack opening plots at $d=60\ \mu\mbox{m}$ (deformation scale=1:50)}
	\label{fig:LPCK}
\end{figure}

The force displacement curves are shown in Figure~\ref{fig:LPanelForce}, generally indicating good agreement with the results obtained by the XFEM \cite{Meschke:01}.  In Figure~\ref{fig:LPanelForce}, the "Original" results are those obtained by the global cracking elements which use higher order elements in the discretized domain from the beginning as used in \cite{Yiming:20,Yiming:21} and only add center node for the cracking elements.  In Figure~\ref{fig:LPanelForce}, it can be found that most results agree with the experimental results obtained by \cite{Winkler:02} and numerical results obtained by \cite{Meschke:01}.  Some deviations can be found in case with Mesh III using "Original" plan, which are caused by the slightly different crack path.  The crack opening plots of all cases at $d=60\ \mu\mbox{m}$ are shown in Figure~\ref{fig:LPCK}, similar to the results obtained by XFEM \cite{Meschke:01}.

\begin{figure}[htbp]
	\centering
	\includegraphics[width=0.99\textwidth]{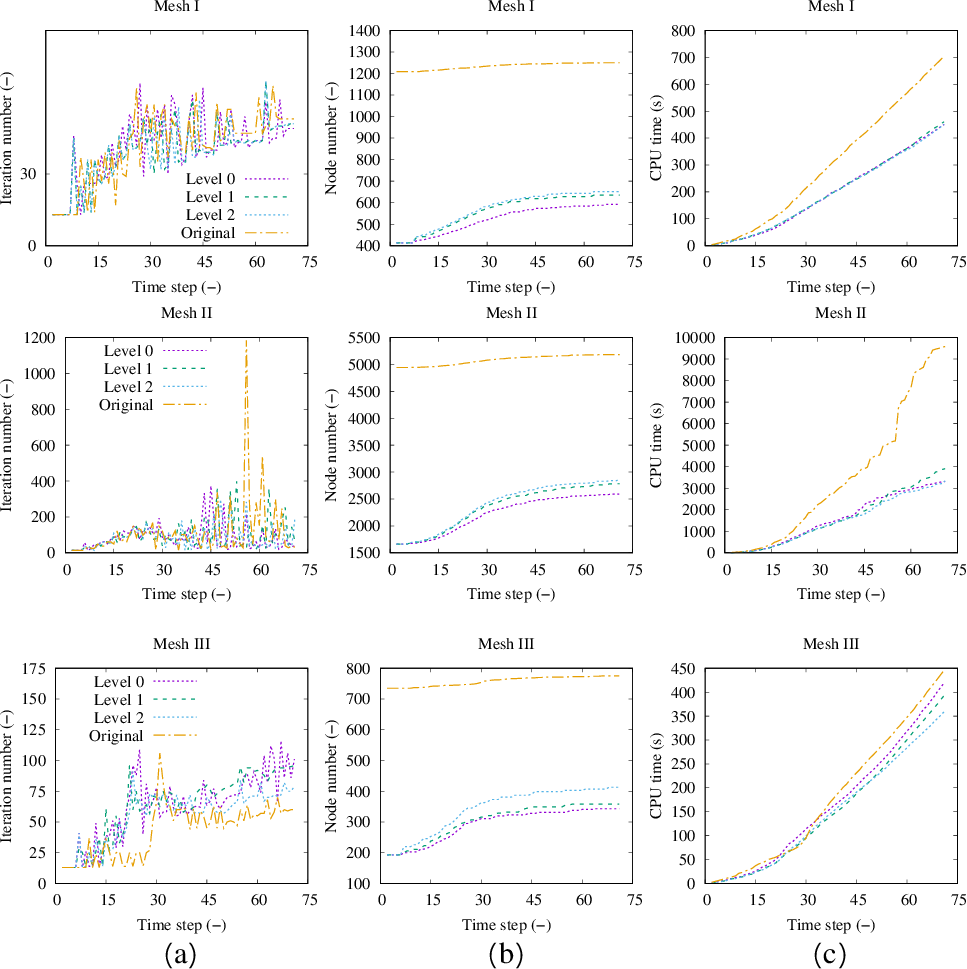}
	\caption{The computing performance: (a) the iteration numbers, (b) the numbers of total nodes, (c) CPU times}
	\label{fig:LPIT}
\end{figure}

For evaluating the computing performance of different strategies, we compare the iteration numbers, the numbers of total nodes and CPU times of all cases, which are shown in Figure~\ref{fig:LPIT}.  Some conclusions can be drawn as
\begin{enumerate}
\item
Considering the iteration number, strategies Level 1 and Level 2 and the original version did not show advantages over Level 0.  Hence, only adding nodes for the cracking elements (Level 0) is a suitable approach;
\item
Adaptive strategies greatly reduce the number of nodes (around 50\%) comparing to the original version of GCEM;
\item
In most cases, great computing efforts can be saved by using the adaptive strategies.  Considering Mesh III, the similar computing time is mainly attribute to the relative coarse mesh.   
\end{enumerate}
Because different adaptive strategies gave similar results, only strategy Level 0 is considered in the remaining parts of this work.

\subsection{Disks with a single initial crack}
Brazilian disk tests with a single initial crack were studied experimentally in \cite{HAERI201420}.  Numerical simulations were conducted by the phase field method \cite{Zhou2019}, the peridynamics theory \cite{ZHOU2016235}, and by the lattice element model \cite{JIANG201741}.  Similar to the GCEM, these methods do not need crack tracking.  However, GCEM is not a damage degree based model but a crack opening model which gives crack openings with relatively coarse meshes.  

\begin{figure}[htbp]
	\centering
	\includegraphics[width=0.97\textwidth]{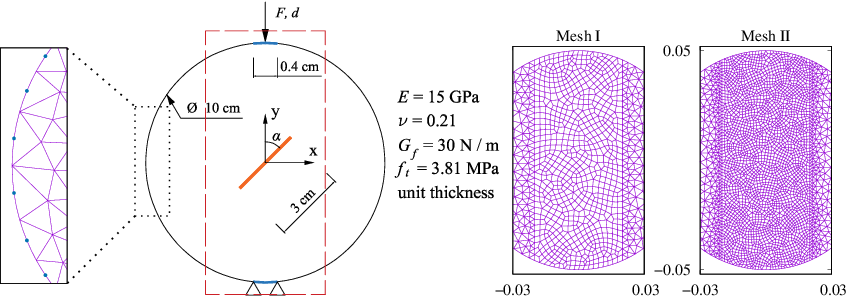}
	\caption{The Disk with a single initial crack: the set-up, material parameters and meshes (unit: cm)}
	\label{fig:DiskModel}
\end{figure}

The model of the tests are shown in Figure~\ref{fig:DiskModel}, with two meshes considered.  Similar to an example provided in \cite{Yiming:23}, the initial crack is not directly modelled but embedded.  In other words, the crack openings and crack directions of the elements on the path of the initial crack are set.  Figure~\ref{fig:DiskModel} shows another advantage of the proposed adding node strategy, though the effects are dispensable in this example.  By adding nodes on the edge of some elements, the curved boundaries can be modelled preciously.

\begin{figure}[htbp]
	\centering
	\includegraphics[width=0.85\textwidth]{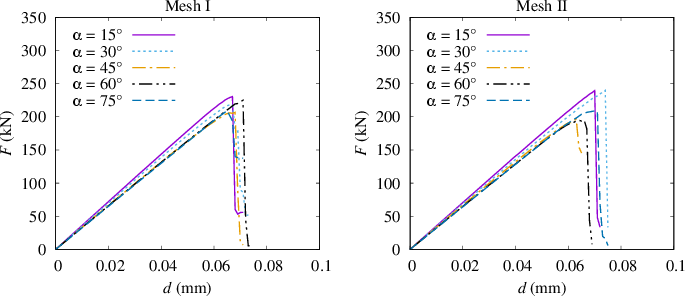}
	\caption{The Disk with a single initial crack: force-displacement curves}
	\label{fig:DiskForce}
\end{figure}

\begin{figure}[htbp]
	\centering
	\includegraphics[width=0.85\textwidth]{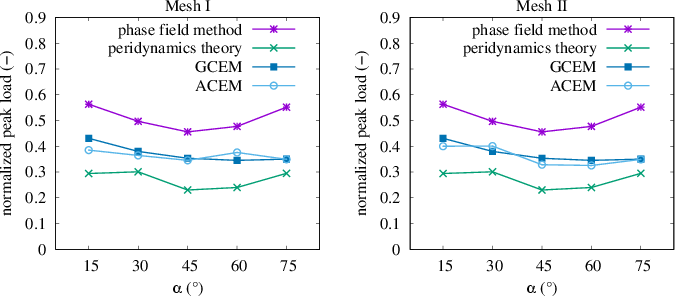}
	\caption{The Disk with a single initial crack: normalized peak loads compared to the results obtained by phase field method \cite{Zhou2019} and peridynamic model \cite{ZHOU2016235} and GCEM \cite{Yiming:21}, where the initial crack was explicit modelled}
	\label{fig:DiskForceNor}
\end{figure}

\begin{figure}[htbp]
	\centering
	\includegraphics[width=0.95\textwidth]{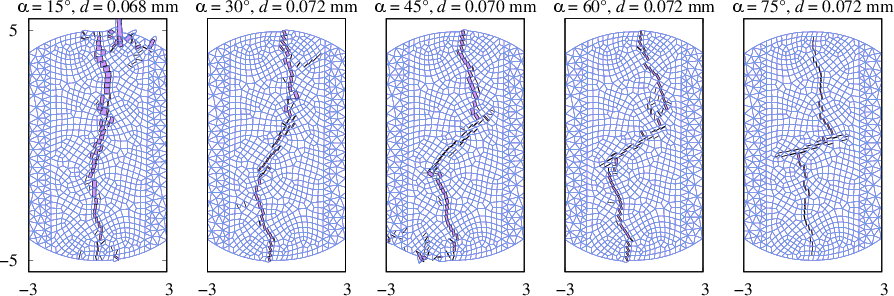}
	\caption{The post peak crack openings plot with Mesh I (deformation scale=1:10)}
	\label{fig:DiskCK}
\end{figure}

\begin{figure}[htbp]
	\centering
	\includegraphics[width=0.95\textwidth]{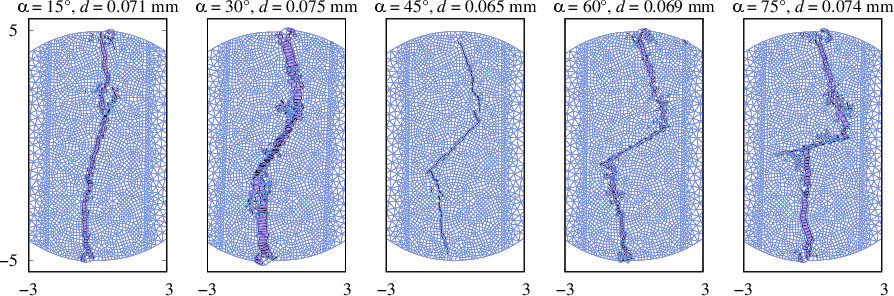}
	\caption{The post peak crack openings plot with Mesh II (deformation scale=1:10)}
	\label{fig:DiskFCK}
\end{figure}

For an intact disk, the analytical peak load per unit thickness is 598.47~\mbox{kN}, which is used to obtain the normalized peak loads.  The force-displacement curves and the normalized peak loads are shown in Figures~\ref{fig:DiskForce}~\ref{fig:DiskForceNor}.  The results indicate that the adaptive crack elements give similar results to those obtained the original global cracking elements \cite{Yiming:21}, which explicitly modelled the initial crack.  Furthermore, the post peak crack openings plots are shown in Figures~\ref{fig:DiskCK} and~\ref{fig:DiskFCK}, agreeing with the experimental results provided in \cite{HAERI201420}.

\subsection{Three-point bending test of a concrete beam}
\begin{figure}[htbp]
	\centering
	\includegraphics[width=0.99\textwidth]{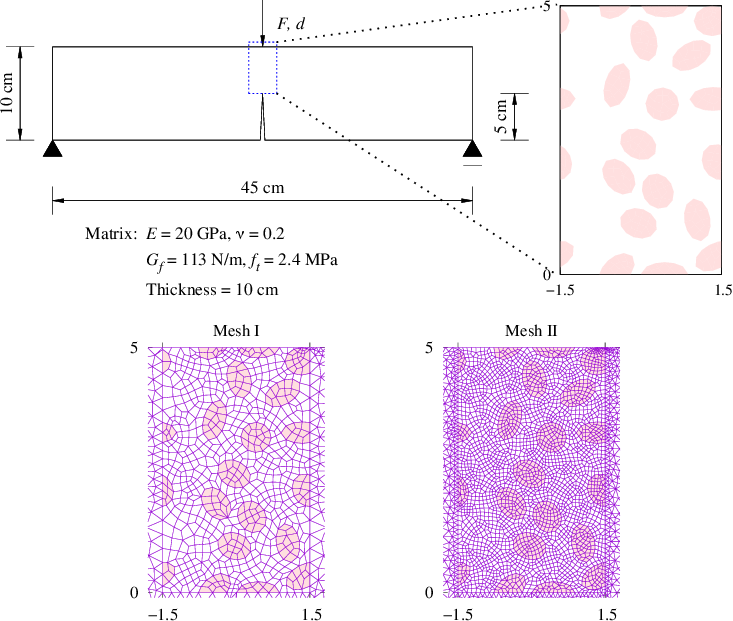}
	\caption{The three-point bending test of a plain concrete beam: the set-up, material parameters (cement paste) and meshes (unit: cm)}
	\label{fig:BeamModel}
\end{figure}

The three-point bending test of a plain concrete beam was experimentally investigated in \cite{Rots:01}.  In this work, we model the middle part of the beam with a matrix-inclusion domain as aggregate-cement paste composite.  The set-up, material parameters (cement paste) and meshes of the test is shown in Figure~\ref{fig:BeamModel}.  Two cases are considered as

\begin{enumerate}
	\item
	Case I: the aggregates have the same material parameters as the cement paste, which was numerically investigated by \cite{Cervera:03,Yiming:11};
	\item
	Case II: the material parameters of the aggregates are: $E=30$~GPa, $\nu=0.2$, $f_t=6$~MPa, $G_f=200$~N/m.  Since the strength of aggregates are much higher than the strength of cements.  The cracks are supposed to bypass the aggregates.  
\end{enumerate}

\begin{figure}[htbp]
	\centering
	\includegraphics[width=0.99\textwidth]{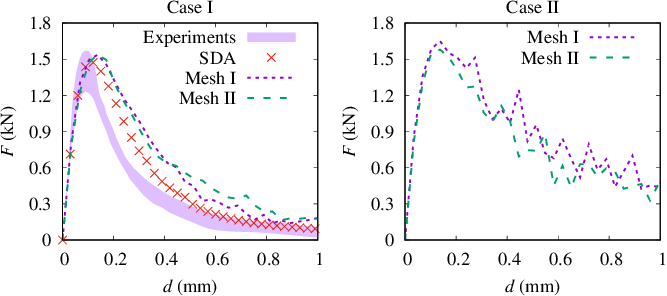}
	\caption{The three-point bending test of a plain concrete beam: force-displacement curves (Considering Case I, the results are compared to the experimental results in \cite{Rots:01} and numerical results obtained by Strong Discontinuity embedded Approach in \cite{Yiming:11} with regular mesh and a prescribed vertical crack)}
	\label{fig:BeamForce}
\end{figure}

\begin{figure}[htbp]
	\centering
	\includegraphics[width=0.99\textwidth]{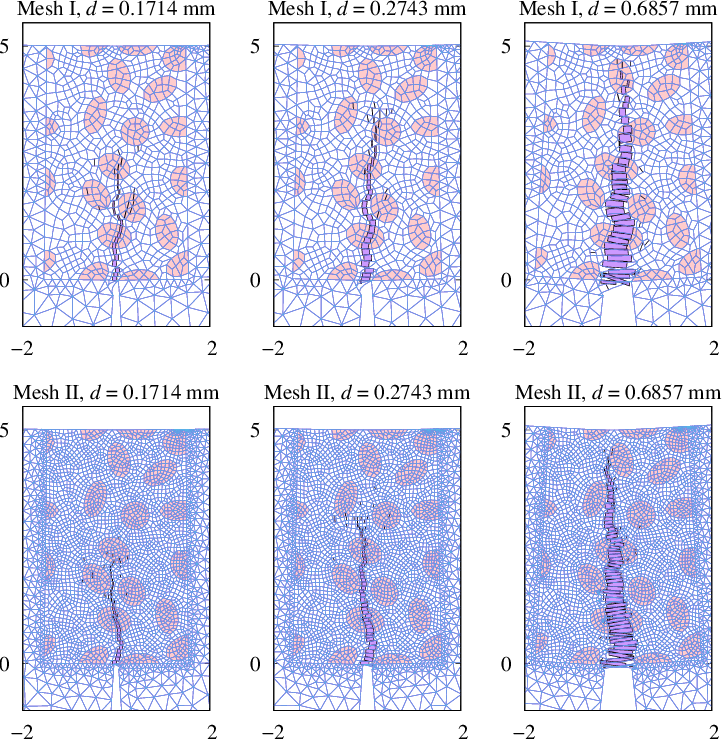}
	\caption{The three-point bending test of a plain concrete beam: Case I crack opening plots at different $d$ (deformation scale=1:20)}
	\label{fig:BeamCase1CK}
\end{figure}

\begin{figure}[htbp]
	\centering
	\includegraphics[width=0.99\textwidth]{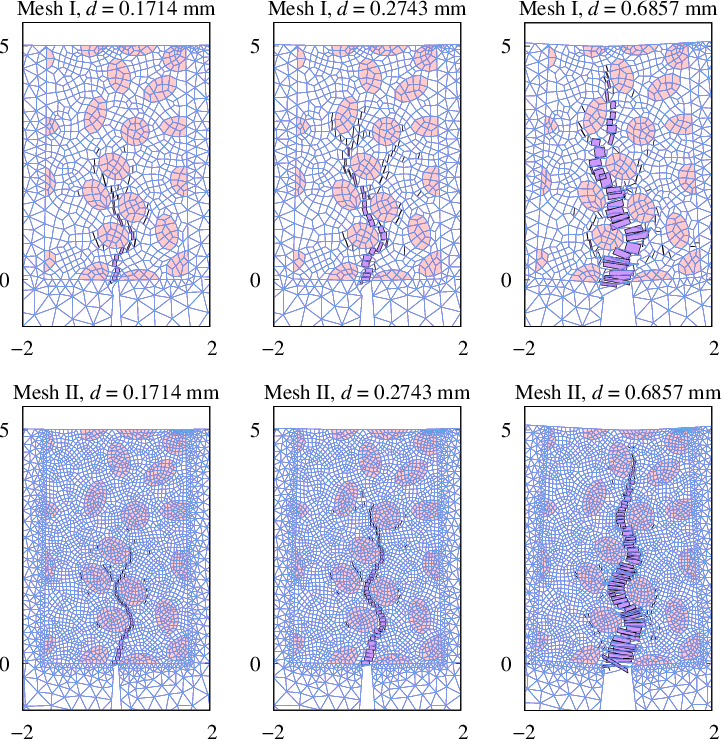}
	\caption{The three-point bending test of a plain concrete beam: Case II crack opening plots at different $d$ (deformation scale=1:20)}
	\label{fig:BeamCase2CK}
\end{figure}

The force-displacement curves are shown in Figure~\ref{fig:BeamForce}.  Considering Case I, our results are generally agreeable with the experimental results in \cite{Rots:01} and the numerical results obtained by the Strong Discontinuity embedded Approach with the standard SOS formulation in \cite{Yiming:11} with regular mesh and a prescribed vertical crack, indicating the reliability of ACEM.  Considering Case II, as expected, the matrix-inclusion model provides higher peak loads and higher post peak residual stress.  Besides, in both cases, Meshes I and II give similar force-displacement results.  The crack openings plots at different $d$ are shown in Figures~\ref{fig:BeamCase1CK} and~\ref{fig:BeamCase2CK}, indicating the crack passing through the aggregates in Case I and bypassing the aggregates in Case II.

\section{Conclusions}
\label{sec:conc}
In this work, aiming at reducing the computing efforts of conventional global cracking elements method which only supports finite elements with nonlinear interpolation of displacement field, a simple hybrid linear and non-linear interpolation finite element is presented.  By using the conventional shape functions of nonlinear elements and modifying the $\mathbf{N}$ vector and $\mathbf{B}$ matrix based on the nodes, a simple computing routine is presented, followed by the Fortran source code.  With this formulation, the edge nodes can be easily added on the original linear elements to upgrade the element to (semi-) nonlinear element or cracking elements.  The numerical examples indicate that our method not only saves great computing time comparing to the original global cracking elements but also shows advantages on precisely modelling curved boundaries.

\section{ACKNOWLEDGMENT}
The authors gratefully acknowledge financial support from the National Natural Science Foundation of China (NSFC) (52178324).



\clearpage
\bibliographystyle{ieeetr}
\bibliography{Reference}







\end{document}